\documentclass[type=submission]{oae}
\usepackage{siunitx}
\usepackage{booktabs}

\usepackage{lineno,hyperref}

\usepackage{amsmath}
\usepackage{booktabs} % For formal tables
\usepackage{indentfirst}
\usepackage{algorithm}
\usepackage{algorithmic}
\usepackage{graphicx}
\usepackage{subfigure}
\usepackage{textcomp}
\usepackage{xcolor}
\usepackage{cleveref}
\usepackage{verbatim}
\usepackage{enumerate}

\documentsetup{
  title = {DCDChain: A Credible Architecture of Digital Copyright Detection Based on Blockchain},
  authors = {
    author1 = {
      name         = {Zhili Chen},
      affiliations = {affil1,affil4},
    },
    author2 = {
      name         = {Yuting Wang},
      affiliations = {affil2},
    },
    author3 = {
      name         = {Tianjiao Ni},
      affiliations = {affil3},
    },
  },
  affiliations = {
    affil1 = {
      Shanghai Key Laboratory of Trustworthy Computing, East China Normal University, Shanghai 200062, China.
    },
    affil2 = {
      School of Computer Science and Technology, Anhui University, Hefei 230601, China.
    },
    affil3 = {
      School of Computer and Information, Anhui Normal University, Wuhu 241003, China.
    },
    affil4 = {
      CCCC Intelligence Transportation Co., Ltd., Tianjin 300210, China.
    },
  },
  correspondence-to = {
    Prof. Zhili Chen, Shanghai Key Laboratory of Trustworthy Computing, East China Normal University, No. 3663, Zhongshan North Road, Shanghai 200062, China. E-mail: zhlchen@sei.ecnu.edu.cn; ORCID: 0000-0002-2231-3652.
  },
  short-author      = {Surname \textit{et al}.},
  journal           = {Journal of Surveillance, \\ Security and Safety},
  short-journal     = {J Surveill Secur Saf},
  how-to-cite       = {},
  received          = {date month year},  % [e.g., 1 Jan 2020]
  first-decision    = {},
  revised           = {},
  accepted          = {},
  published         = {},
  academic-editor   = {},
  copy-editor       = {},
  production-editor = {},
  doi               = {10.20517/jsss.xxxx.xx},
  year              = {Year},
  volume            = {Volume},
  end-page          = {Number},
}

\usepackage{hyperref}

\begin{document}

\maketitle

\begin{abstract}
  \hint{Suggestions: No more than 250 words. No citations. Define abbreviations at their first mention.}

  \paragraph{Aim:} Copyright detection is an effective method to prevent piracy. However, untrustworthy detection parties may lead to falsified detection results. Due to its credibility and tamper resistance, blockchain has been applied to copyright protection. Previous works mainly utilized blockchain for reliable copyright information storage or copyrighted digital media trading. As far as we know, the problem of credible copyright detection has not been addressed.

  \paragraph{Methods:} In this paper, we propose a credible copyright detection architecture based on the blockchain, called DCDChain. In this architecture, the detection agency first detects copyrights off the chain, then uploads the detection records to the blockchain. Since data on the blockchain are publicly accessible, media providers can verify the correctness of the copyright detection, and appeal to a smart contract if there is any dissent. The smart contract then arbitrates the disputes by verifying the correctness of detection on the chain. 

  \paragraph{Results:} The detect-verify-and-arbitrate mechanism guarantees the credibility of copyright detection. Security analysis and experimental simulations show that the digital copyright detection architecture is credible, secure and efficient.

  \paragraph{Conclusion:} The proposed credible copyright detection scheme is highly important for copyright protection. The future work is to improve the scheme by designing more effective locality sensitive hash algorithms for various digital media.

  \paragraph{Keywords:} Digital copyright detection, blockchain, credibility, tamper resistance

  \hint{Please suggest 3-8 keywords which can be used for describing the content of the manuscript and will enable the full text of the manuscript to be searchable online.}
\end{abstract}

\section{1. Introduction}

With the popularization of digital technologies, there are more and more ways to copy and transmit digital media \cite{sanchez2016sci}. Since many netizens lack a good awareness of copyright problems, copyrights are often not fully protected, and the benefits of owners are damaged frequently. To alleviate the copyright problems, researches on copyright protection are crucial.

Recently, a large number of papers have proposed measures to prevent piracy \cite{sabeeh2021plagiarism,cheers2021academic,chang2021using,gupta2016study}. A common approach is plagiarism detection which is mainly to determine whether the media has copied or plagiarized content. Vani et al. \cite{vani2018integrating} proposed that texts are segmented and then tested for plagiarism detection using the vector space model\cite{ekbal2012plagiarism}. Change et al.\cite{chang1998rime} presented a content-based image plagiarism detection for the first time. Zhou et al. \cite{zhou2016effective} proposed an efficient image copy detection to resist arbitrary rotations. Wary et al.\cite{wary2018review} reviewed the research progress of piracy detection in video. These researches depend on the detection agencies or detection tools which maybe not fully credible. If the detection agencies are corrupted or the detection tools \cite{sharma2015plagiarism} (such as Parikshak, Copy Scape) crashes, the loss will be immeasurable. The research \cite{weber2013plagiarism} also proved that the detection tools may not get correct detection results. So plagiarism detection tools should not be blindly trusted.

Due to the credibility and tamper resistance nature, blockchain has received extensive attention in the copyright protection field \cite{wang2022image,wang2021algorithmic,nizamuddin2019blockchain,cheng2018blockchain,savelyev2018copyright,fujimura2015bright,xu2017design}. However, it is only used as a trading platform\cite{bhowmik2017multimedia} to earn copyright fees for copyright owners \cite{nizamuddin2019blockchain,smith2018transacting}, or to reliably store media copyright records \cite{cheng2018blockchain,savelyev2018copyright,fujimura2015bright,xu2017design}. In \cite{ma2018blockchain,meng2018design}, an image is embedded with digital watermarks, and its hash value is stored on the blockchain, ensuring that the existence proof of the exact copyrighted image cannot be forged. As far as we know, there is a small amount of blockchain-based work to address the issue of copyright detection, where media content may be completely or partially pirated. The smart contracts that use perceptual hashing and ethereum automatically detect and reject tampered images that are perceptually similar to those already on the market \cite{mehta2019decentralised}. But the design is only for the image, the lack of different media piracy detection consideration. There is no mention of the full design process of the smart contract and the security considerations for the possibility of copyright theft of the image itself.

%Research regarding that the blockchain in copyright detection is currently still in its infancy \cite{savelyev2018copyright,fujimura2015bright,xu2017design}.  It provides a way for users to verify the authenticity of copyright through mobile phone scanning or website inquiries \cite{cheng2018blockchain}. Media usage rights as a commodity, copyright owners, publishers and buyers trading on the blockchain platform \cite{nizamuddin2019blockchain}. But there is no copyright detection consideration.

In this paper, we propose a credible copyright detection architecture. Specifically, to ensure reliable detection results, the detection agency detects the copyrights of media locally, and stores the results on blockchain. Then due to the public accessibility of blockchain, the media providers can verify the correctness of the detection results locally, and appeal to a smart contract for arbitration if there is any dissent. The smart contract then arbitrates the disputes by the detection results on the chain. Furthermore, financial incentives are also used to stimulate both the detection agency and the media providers. Our main contributions can be summarized as follows:
\begin{itemize}
\item To our best knowledge, we first propose a common copyright detection scheme based on blockchain, which ensures credible copyright detection and is suitable for various media.
\item We perform credibile, accurate and efficient copyright detection based on detect-verify-and-arbitrate mechanism and by combining secure hashing and locality sensitive hashing innovatively.
\item We conduct experiments in four different corpora and use regression models to determine the relationship between hamming distances and similarities of digital media. We implement smart contract on both Rinkeby testnet and local Ganache, and compare them in term of time. Experimental results show that our scheme is feasible and efficient.
\end{itemize}

The rest of the paper is organized as follows. Section 2 gives technical preliminaries. Section 3 describes the situation analysis and design targets. In Section 4, we present the overview and the detailed design of our scheme. Section 5 is evaluation and discussion. Finally, Section 6 concludes our work.

\section{2. Methods}\label{sec:methods}

\subsection{2.1 Technical Preliminaries}\label{sec:preliminaries}

In this section, we introduce some technologies related to our architecture.

\subsection{2.1.1 Blockchain}

Blockchain \cite{nakamoto2008bitcoin,zheng2018blockchain} is a technology of distributed ledgers. Except for the genesis block, in a blockchain, each block contains the hash value of the previous block and a timestamp indicating write time of data, so any data on blockchain are tamper-resistant. Peer nodes encapsulate transactions into a block and link it to the current longest main blockchain. Every transaction can be traced through this chain structure. Also, every transaction in the block contains the initiator's digital signature to ensure authenticity and legality. The tamper resistance, traceability, and authenticity of the blockchain are useful for copyright detection.

Smart contracts are event-driven computer programs that are deployed in the blockchain. They are regarded as special transactions that are easily traced and cannot be tampered or reversed. Once the predefined conditions are met, the code on the contract can be executed automatically without the external authorization. 

\subsection{2.1.2 Hash algorithm}

Our scheme is based on two kinds of hash algorithms:  secure hash and locality sensitive hash. The marked hash value of digital media is implemented by the secure hash algorithms.

%\begin{mydef}[Secure hash algorithm]
\textbf{Definition 1 (Secure Hash)}
A secure hash algorithm is a  one-way function $H$ which accepts an arbitrarily large input $x$, and produces a small fixed-size output $h$: $h=H(x)$.
It is a function with the following basic properties.
\begin{itemize}
\item Collision resistance: it should be difficult to find two distinct inputs $x$, $x^\prime$ such that $H(x)=H(x^\prime)$.
\item Determinism: it is a data conversion function that maps inputs (raw data) to fixed-length outputs (indexes).
\end{itemize}
%\end{mydef}

We calculate the media similarity based on the locality sensitive hash algorithm.

\textbf{Definition 2 (Locality Sensitive Hash)}\cite{datar2004locality}
For any two points $x$, $x^\prime$ in space $S$ and the distance $D(x, x^\prime)$ of the two points,  such as Euclidean distance, Manhattan distance, etc., if the function $H=\{h: S \rightarrow U\}$ meets the following two conditions:
\begin{itemize}
  \item $D(x, x^\prime)\leq r, Pr_{H}\lbrack h(x)=h(x^\prime)\rbrack\geq p_1$
  \item $D(x, x^\prime)>r(1+\varepsilon),\;Pr_{H}\lbrack h(x)=h(x^\prime)\rbrack\leq p_2$
\end{itemize}
where $\varepsilon, r$ are positive integers and $p_1 > p_2$, then $H=\{h:S\rightarrow U\}$ is a locality sensitive hash algorithm.

\subsection{2.1.3 IPFS}

The Interstellar File System (IPFS) is a peer-to-peer distributed file system that aims to replace HTTP. IPFS combines many good ideas in the current file systems. BitSwap Protocol solves the problem of file sharing. Merkle DAG is another important part of IPFS because it guarantees that IPFS has useful functions such as content addressing, tamper resistance and deduplication. It replaces the domain-based address with a content-based address and the user can find the file stored in it by the address hash value.

\subsection{2.2 DESIGN TARGETS}\label{sec:targets}

In the real copyright detection process, there may be opacity of the detection process and the evaluation criteria. For example, for a paper, different paper examination tools may have different reproduction ratios. Thus, when the same paper has different test results, we do not know which one can be believed. Moreover, the tester may maliciously provide incorrect test results. The mistakes of results are not easy to find. Even if we suspect the test results, we cannot submit our comments. 

Therefore, our design achieves the following three targets.

\textbf{Tamper resistance:} All copyright data should be stored tamper-resistantly, such that anyone cannot change the copyright data maliciously. 

\textbf{Credibility:} The copyright detection results should be entirely reliable. The detection process should be carried out in a credible way, and the detection results should be publicly verifiable.

\textbf{Efficiency:} The copyright detection should be sufficiently efficient. It should be fast enough to detect the copyrights of digital media products.

%Since the main goal is to protect the original media and get the correct detection results, we assume that original media is encrypted and no one is able to access original media without decryption key.

\section{2.3 DCDCHAIN design}\label{sec:overview}

In this section, we propose a credible architecture of digital copyright detection based on blockchian, called DCDChain, and present the details of the design.

\subsection{2.3.1 Overview}

\begin{figure}
\centering
\includegraphics[scale=0.45]{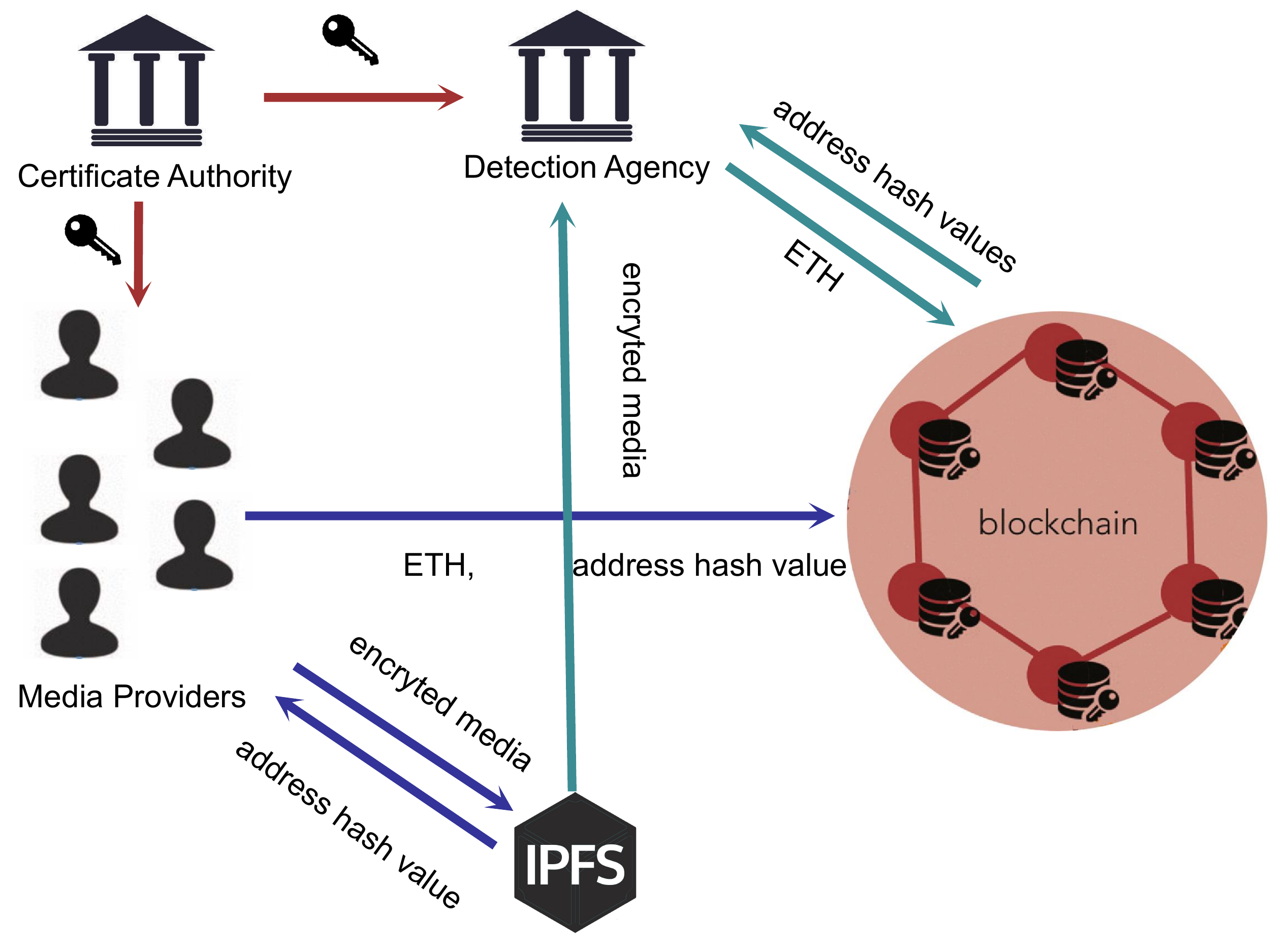}
\caption{Overview of the architecture}
\label{fig:label1}
\end{figure}

We adopt the architecture as illustrated in Figure~\ref{fig:label1}. There are five entities involved, namely, detection agency (DA), media providers (MPs), certificate authority (CA), Blockchain, and inter planetary file system (IPFS). 

The DA aims to earn service fees by providing copyright detection services, while MPs want to obtain the sevices.  The detection services include the detection of pirated media and the announcement of results according to encrypted media. DA's account and identity are public, and the detection algorithm are also made public, which is key for MPs' verification of detection results. CA authenticates parties in the system by providing digital certificates for their public keys.  The blockchain holds all the copyright-protected digital media. Furthermore, to automatically and reliably arbitrate the detection results, DA deploys a smart contract with arbitration and transfer functions on the blockchain. The arbitration function includes two aspects: one is to determine whether hash values provided by MPs correspond to legal digital media; the other is to check whether detection results of doubtful content satisfy the piracy confirmation rule. 

The work flow of DCDChain is as follows.
\begin{enumerate}[(1)]
\item MPs store complete digital media on the IPFS and get back their corresponding adress hash values, which in turn are stored on the blockchain. Digital media can be encrypted before storage if necessary. Note that it is unrealistic to store digital media directly on the blockchain due to large sizes. We adopt a more practical approach \cite{cheng2018blockchain,zhaofeng2018new} that stores hash values uniquely indexing digital media. These hash values are calculated by the SHA256, so once the digital media have any changes, they change significantly. 

\item When MPs want to request copyright detection services, they send the blockchain the address hash values indexing digital media to detect, and pay it service fees. Note that these hash values and service fees are in fact sent to the smart contract deployed on the blockchain. 

\item On receiving the detection requst, the smart contract stores the service fees, and sends the address hash values to DA. Since the detection is computation-intensive, we let DA do it locally off the blockchain and only send the detection results to the smart contract.

\item On receiving the address hash values, DA pays the smart contract a deposit, conducts the copyright detection, and then returns the detection results to the smart contract. Since the direct copyright detection with media contents is very computation-intensive, we adopt the locality sensitive hashing (LSH) to calculate the similarities of digital media, and improve the detection efficiency significantly.

\item MPs verify the detection results. If they suspect the results, they are able to issue a doubt to the smart contract, which then enters the arbitration procedure. 

\item
Once arbitrating that the detection results are wrong, the smart contract automatically sends the deposit to MPs. Otherwise, if arbitrating the detection results are right or there is no arbitration, the smart contract sends the service fees to DA.

\end{enumerate}

The main notations of the paper are shown in Table$\ref{table1}$.

\begin{table}
\centering
\caption{NOTATIONS}
\label{table1}
\begin{tabular}{|c|c|}
\hline
Notations & Descriptions \\

\hline
$m$   &  Medium to be detected \\
\hline
${\emph{hashID}}$ & The SHA256 hash value of $m$ \\
\hline
$\emph{lshv}$  &  The LSH hash value of $m$ \\

\hline
$\emph{lm}_j$ & The $j$-th legal media\\
\hline
$\emph{hashID}_{j}$ & The SHA256 hash value of $\emph{lm}_j$\\
\hline
$\emph{lshv}_{j}$ & The LSH hash value of $\emph{lm}_j$\\

\hline
$f_{i}$ & The service fee\\
\hline
$f_{\emph{DA}}$ & The deposit of DA\\

\hline
$L$  & Hamming distance  \\
\hline
$\theta$  &  Threshold of hamming distance \\

\hline
\end{tabular}
\end{table}

\subsection{2.3.2 Detail Design}

We apply a hybrid approach that incorporates blockchain, hash algorithms and IPFS into our design. It allows the detection agent (DA) to calculate the similarities of media credibly, while the media providers (MPs) and the public are able to trace copyright information. As the trusted party, CA authenticates the identities of DA and MPs, and issues the digital certificates including public keys $\emph{PK}_{\emph{DA}},\emph{PK}_{i}$ to them, respectively. The corresponding private keys $\emph{SK}_{\emph{DA}}$, $\emph{SK}_{i}$ are kept secret individually.

The proposed credible and tamper resistance media transaction architecture based on blockchain model consists of two stages: preliminary stage and detect-verify-and-arbitrate stage. The proposed architecture is depicted in Figure~\ref{fig:label1}.

\textbf{(1) Preliminary Stage}

There are three steps in the preliminary stage. First, the media provider $\emph{MP}_{i}$ and the detection agent $\emph{DA}$ authenticate each other with digital certificates released by the certificate authority (CA). Second, the original media file $m$ is encrypted with mix encryption to ensure that only the detection agent $\emph{DA}$ can decrypt it. Third, the encrypted media file is uploaded to IPFS and its address hash is handed to the smart contract and also to detection agent $\emph{DA}$. At the same time, the media provider $\emph{MP}_i$ commits a testing service fee and the detection agent $\emph{DA}$ commits a security deposit to the smart contract. The detailed description is as follows.

\textbf{Step 1}: The media provide $\emph{MP}_{i}$ and the detection agent $\emph{DA}$ verify each other's identity by digital certificates as follow:
\begin{enumerate}[(1)]
\item $\emph{MP}_i$ sends its digital certificate $C_{\emph{MP}_i}$ to $\emph{DA}$, and $\emph{DA}$ sends its digital certificate $C_{\emph{DA}}$ to $\emph{MP}_i$.

\item Both $\emph{MP}_i$ and $\emph{DA}$ verify each other's digital certificates with the public key of the certificate authority.

\item If either $\emph{MP}_i$ or $\emph{DA}$ finds an invalid digital certificate, it abort the protocol. Otherwise, both of them extract the identity and public key of each other from the digital certificate.
\end{enumerate}

\textbf{Step 2}: The media provider $\emph{MP}_i$ encrypts the media file $m$ with the mix encryption. Specifically, $\emph{MP}_{i}$ encrypts the media by asymmetric encryption $E^a$ and symmetric encryption $E^s$ as \cref{eq:mix-encrypt} and gets the cipertext $S_{i}^m$.
\begin{equation}\label{eq:mix-encrypt}
S_{i}^m=(E^a_{PK_{DA}}(k_i),E^s_{k_i}(m))
\end{equation}
which can be decrypted only by the detection agent $\emph{DA}$ with its private key $\emph{SK}_{\emph{DA}}$ as \cref{eq:mix-decrypt}.
\begin{equation}\label{eq:mix-decrypt}
\begin{aligned}
k_i &=D^a_{\emph{SK}_{\emph{DA}}} (E^a_{\emph{PK}_{\emph{DA}}}(k_i))\\
m &= D^s_{k_i} (E^s_{k_i}(m))
\end{aligned}
\end{equation}

\textbf{Step 3}: The media provider $\emph{MP}_{i}$ uploads $S_{i}^m$ to IPFS and gets back the address hash value $Q_{i}^m$. $\emph{MP}_{i}$ then stores $Q_{i}^m$ in the smart contract denoted by $\emph{SC}$, which has been deployed on the blockchain by the detection agent $\emph{DA}$. At the same time, $\emph{MP}_{i}$ sends a service fee $f_{i}$ and DA sends a deposit $f_{\emph{DA}}$ to $\emph{SC}$. The deposit is used to penalize the DA who perform dishonestly, while the service fee is used to pay for the service. 

The preliminary stage above guarantees that the corresponding $S_{i}^m$ can be found through $Q_{i}^m$ and only $\emph{DA}$ can decrypt it to get the original content of $m$. However, there is only read permission for DA, and it cannot tamper with the contents of $m$. The existing record of $Q_{i}^m$ in IPFS can be used as an important proof of existence for copyright protection of $m$.

\textbf{(2) Detect-Verify-and-Arbitrate Stage}

\begin{figure}
\centering
\includegraphics[scale=0.45]{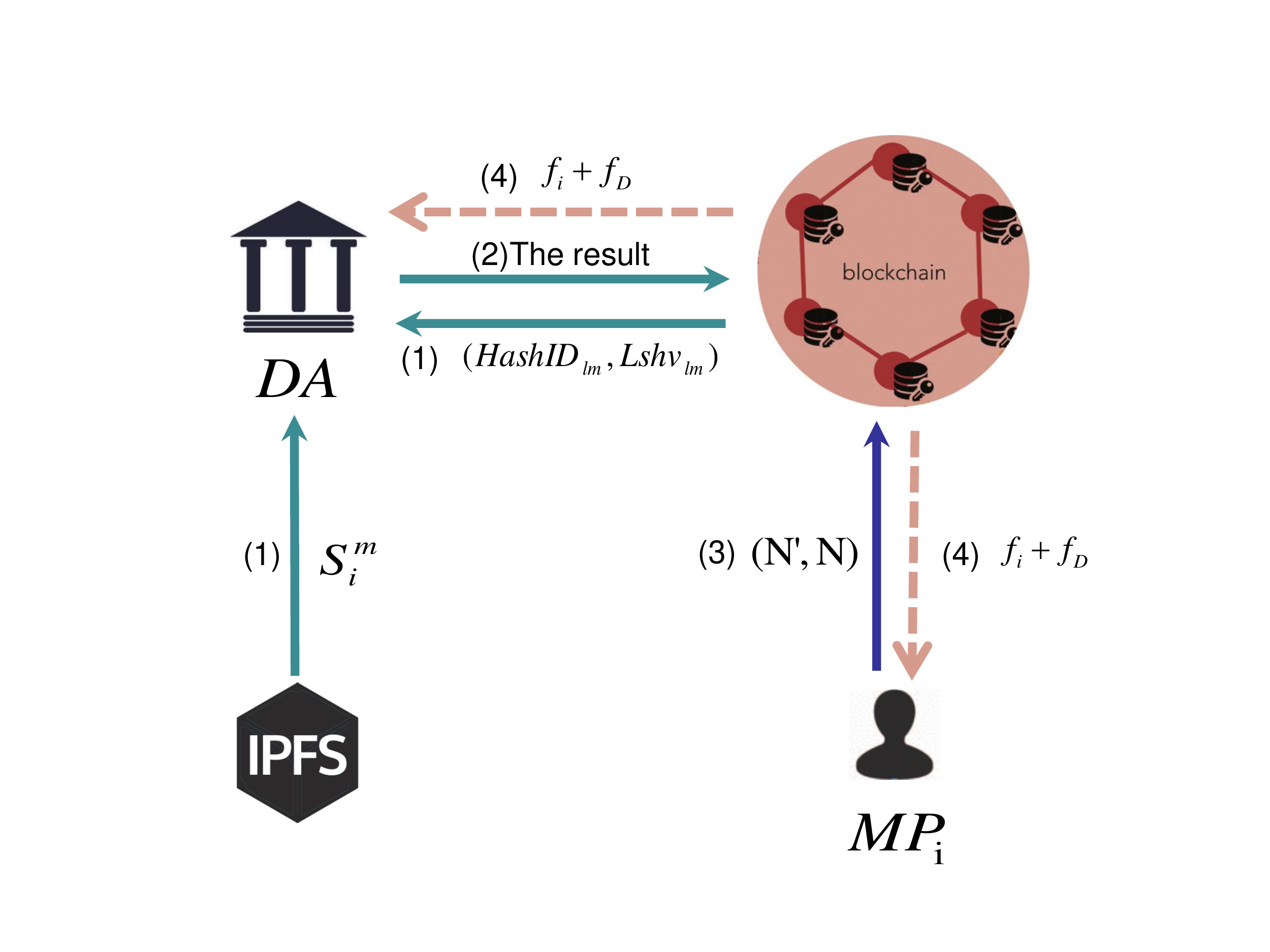}
\caption{Detect-verify-and-arbitrate workflow}
\label{fig:label01}
\end{figure}

In this stage, the detection agent $\emph{DA}$ first obtains the content of the media file $m$, calculates its hash values, and detects if this media file is pirated or not. This detection result will be verified by media providers $\emph{MP}_i$, and finally arbitrated by the smart contract $\emph{SC}$. The work flow is shown as in Fig.~\ref{fig:label01}.

\textbf{Detect:} The detection agent $\emph{DA}$ downloads $S_{i}^m$ from IPFS by $Q_{i}^m$, and then decrypts $S_{i}^m$ by $\emph{SK}_{\emph{DA}}$ to get $m$.
DA calculates $\emph{hashID}$ and $\emph{lshv}$ by SHA256 and LSH algorithms, respectively. At the same time, it downloads all hash values of legal media $\emph{HashID}_{\emph{lm}}$ and $\emph{Lshv}_{\emph{lm}}$ from the blockchain, where
$\emph{HashID}_{\emph{lm}}=\{\emph{hashID}_{1},...,\emph{hashID}_{j},...\}$, $\emph{Lshv}_{\emph{lm}}=\{\emph{lshv}_{1},...,\emph{lshv}_{j},...\}$. Then,

\begin{enumerate}[(1)]
\item Match $\emph{hashID}$ in $\emph{HashID}_{\emph{lm}}$, and if $\emph{hashID} = \emph{hashID}_{j}$ for some $j$, $m$ is complete pirated media.
\item If $\emph{hashID} \notin \emph{HashID}_{\emph{lm}}$, calculate hamming distance $L(\emph{lshv},\emph{lshv}_{j})$ for all $\emph{lshv}_j \in \emph{Lshv}_{\emph{lm}}$. If $L(\emph{lshv},\emph{lshv}_{j})<=\theta$ exists, $m$ is partial pirated media.
\end{enumerate}

Algorithm~\ref{algorithm1} gives the judgment rules for pirated media.

\renewcommand{\algorithmicrequire}{\textbf{Input:}}  %Use Input in the format of Algorithm
\renewcommand{\algorithmicensure}{\textbf{Output:}}  %UseOutput in the format of Algorithm
\begin{algorithm}[htb]
  \caption{The judgment rules for pirated media}
  \label{algorithm1}
  \begin{algorithmic}[1]
  \REQUIRE $m$
  \ENSURE the result
  \STATE DA calculates $\emph{hashID}, \emph{lshv}$.
  \STATE DA downloads $\emph{HashID}_{\emph{lm}},\emph{Lshv}_{\emph{lm}}$.
  \IF{$\emph{hashID} \in \emph{HashID}_{\emph{lm}}$}
  \STATE Complete piracy.
  \ELSE
  \IF{$L(\emph{lshv},\emph{lshv}_{j}) \le \theta$ for some $\emph{lshv}_j \in \emph{Lshv}_{\emph{lm}}$}
  \STATE Partial piracy.
  \ELSE
  \STATE Legitimate media.
  \ENDIF
  \ENDIF
  \STATE DA sends the result to $\emph{SC}$.
  \end{algorithmic}
\end{algorithm}

Note that in practical implementations, the detection agent $\emph{DA}$ can achieve more efficient searches as follows. $\emph{DA}$ maintains a local database exactly the same as that of the blockchain but with more efficient search structures, and detects media files with the local database and uploads detection results to the blockchain accordingly. $\emph{DA}$ can backup all detection results to its local database and only need to update the database when there is a new detection result.

The pirated results include two types: partial piracy and complete piracy. DA uploads the result, $[N,\emph{lshv},\emph{QM}]$ or $[N,\emph{hashID},\emph{QM}]$, to $\emph{SC}$. Among them, $N$ stands for the serial number of legitimate media on the blockchain detected by DA, and $\emph{QM}$ represents the corresponding address hash value of the detected media on IPFS, which can be indexed to the original encrypted media. Both hash values correspond to each other, forming an tamper-resistant record on the blockchain. If there is a conflict, it can be used as a proof of detection existence. In addition to the above, if $m$ is detected as a legitimate media, $\emph{DA}$ uploads $\{N',\emph{hashID},\emph{lshv},\emph{QM}\}$ to $\emph{SC}$, where $N'$ is a newly stored legitimate media serial number. In order to ensure the correctness of the test results, next we need to verify-and-arbitrate the detect results.

\textbf{Verify-and-Arbitrate:} In order to make the detection result real and trustworthy, we use the economic incentive mechanism of blockchain in our design, and the party with the correct detection result will get an amount of digital cash. The source of the bonus is the deposit $f_{\emph{DA}}$ of $\emph{DA}$, and the service fee $f_{i}$ of $\emph{MP}_{i}$.

The smart contract $\emph{SC}$ is an important part of the blockchain. Since smart contracts are trusted and auto-executable, we let $\emph{SC}$ take on the responsibility of validation and arbitration. If the detect result is partial piracy or complete piracy, then after DA uploads the result, SC obtains $\emph{hashID}_{j}$ or $\emph{lshv}_{j}$ of the legitimate media by the serial number $N$. $\emph{SC}$ uses the verification mechanism to directly verify the hash values to detect according to the judgment rule. After successful verification, $\emph{SC}$ transfers the service fee $f_{i}$ of this detection and the deposit $f_{\emph{DA}}$ to DA.

\begin{figure}
\centering
\includegraphics[scale=0.5]{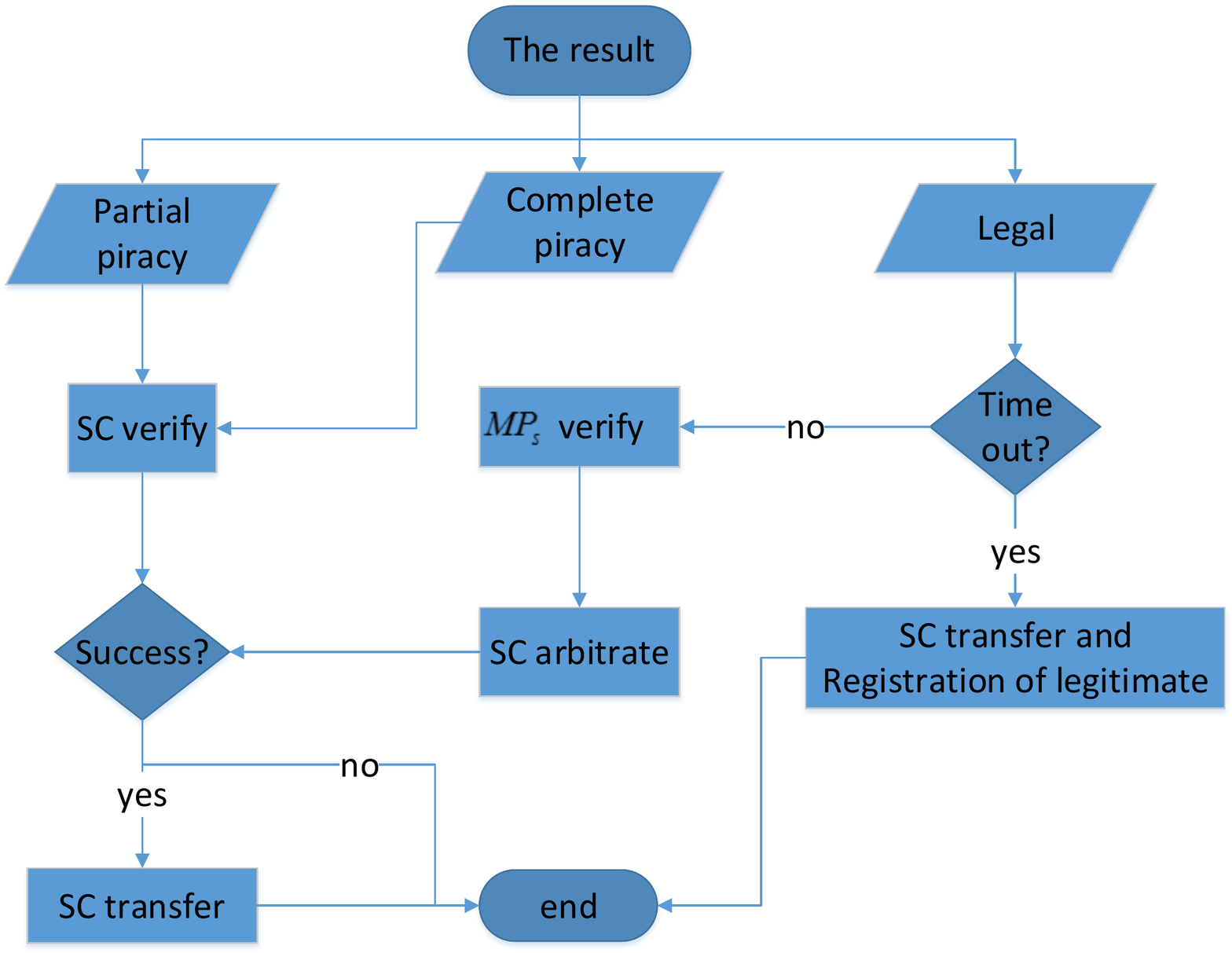}
\caption{Verify-and-arbitrate workflow}
\label{fig:label02}
\end{figure}

If the detect result is illegal, $\emph{MP}_{i}$ can verify the correctness of the copyright detection and cause $\emph{SC}$ to arbitrate the result, as shown in Figure~\ref{fig:label02}. If someone in MPs challenges the detect result, it can initiate arbitration claim to $SC$ by providing an evidence $\{N',N\}$. Specifically, $\emph{MP}_{i}$ verifies locally the detection results on the blockchain, and then uploades related evidences to the blockchain if there are disagreements. $\emph{MP}_{i}$ can back up all the hash values to its own database, and only needs to update the database when new detect results are available. Within time $T$, $\emph{SC}$ obtains the newly detected hash values $\{\emph{hashID}',\emph{lshv}'\}$ and the hash values of the legal media $\{\emph{hashID},\emph{lshv}\}$ by the evidence $\{N',N\}$, respectively. The $SC$ conducts arbitration according to the piracy judgment rules. If $\emph{MP}_{i}$ challenges successfully, $\emph{SC}$ rewards $\emph{MP}_{i}$ with deposit $f_{\emph{DA}}$, and thus both $f_{DA}$ and $f_{i}$ will be transferred to $MP_{i}$ together. If the chanllenge is unsuccessful, $\emph{SC}$ indicates that $\emph{MP}_{i}$ has verified incorrectly, and $SC$ does nothing. If the time $T$ is exceeded, it is regarded that all MPs agree with the detect result of DA, and $\emph{SC}$ will transfer the service fees to DA. 

\subsection{2.3.3 Security Analysis}

We rely on blockchain to achieve credibility and tamper-resistance. We assume a sufficiently large and untrusted peer-to-peer network. In addition, we assume that there are already enough legitimate address hash values stored on the blockchain, which is an accumulation process of the legitimate media. We now show how our design is secure against adversaries.

First, in the design, adversaries cannot tamper a user's data, and no one except the DA can decrypt data. The decentralized nature of blockchain ensures that adversaries cannot disrupt the network and tamper with others' data. Meantime, the security of the digital signature scheme ensures that adversaries cannot impersonate a user and tamper its data. Moreover, the adversaries cannot obtain any information of others from the public ledger because only the hash pointer is stored on it. Actually, the encrypted media $S_{i}^m$ are stored on the IPFS in a tamper-resistant manner, and only the DA who possesses the $\emph{SK}_{\emph{DA}}$ can decrypt it.

Second, media providers $\emph{MP}_{i}$ cannot get detection services without paying service fees to the detection agent. For a correct detection, the smart contract $\emph{SC}$ will verify successfully, and then arbitrate that the detection result is correct if necessary, and finally pay service fees to DA. This is executed automatically, and no one can prevent it due to the credibility of the blockchain. 

Third, the detection agent $\emph{DA}$ cannot get the service fees without providing correct detection. For an incorrect detection, the smart contract $\emph{SC}$ will verify unsuccessfully, and pay the deposit of $\emph{DA}$ to $\emph{MP}_{i}$. Moreover, the media providers $\emph{MP}_{i}$ can appeal to $\emph{SC}$ by providing evidences, causing $\emph{SC}$ to arbitrate that the detection result is incorrect. 

In a word, the cridibility and tamper-resistance properties of the blockchain and the IPFS ensure the security of our design.

\section{3. Resuts}\label{sec:evaluation}

In this section, we conduct experimental simulations and evaluate the performance of DCDChain.
All the experiments are performed on a machine with Intel(R) Core(TM) i5-6200U CPU @ 3.20 GHz and 8GB RAM, running Windows 10. The software used in the experiment includes Pycharm, Atom and Ethereum wallet (MetaMask).

\subsection{5.1 Experimental simulations}

DCDChain is a general architecture, but in this part, we use text for simulation.

\subsubsection{5.1.1 Detection Section}

In our copyright detection, we use hamming distances between LSH hash values to represent media similaries. Therefore, we first do experiments to verify the relationship between hamming distances and media similaries.

There are several classical categories of LSH hash. The common categories in the text are k-shingle, simhash and minhash. The common categories in the image are perceptual hashing, average hash, and different hash algorithms. Our design aims to detect piracy in a large number of text, so we choose simhash for simulation.

\begin{figure}
  \centering
  \subfigure[cda]{
    %\label{fig:subfig:a} %% label for first subfigure
    \includegraphics[width=2.2in]{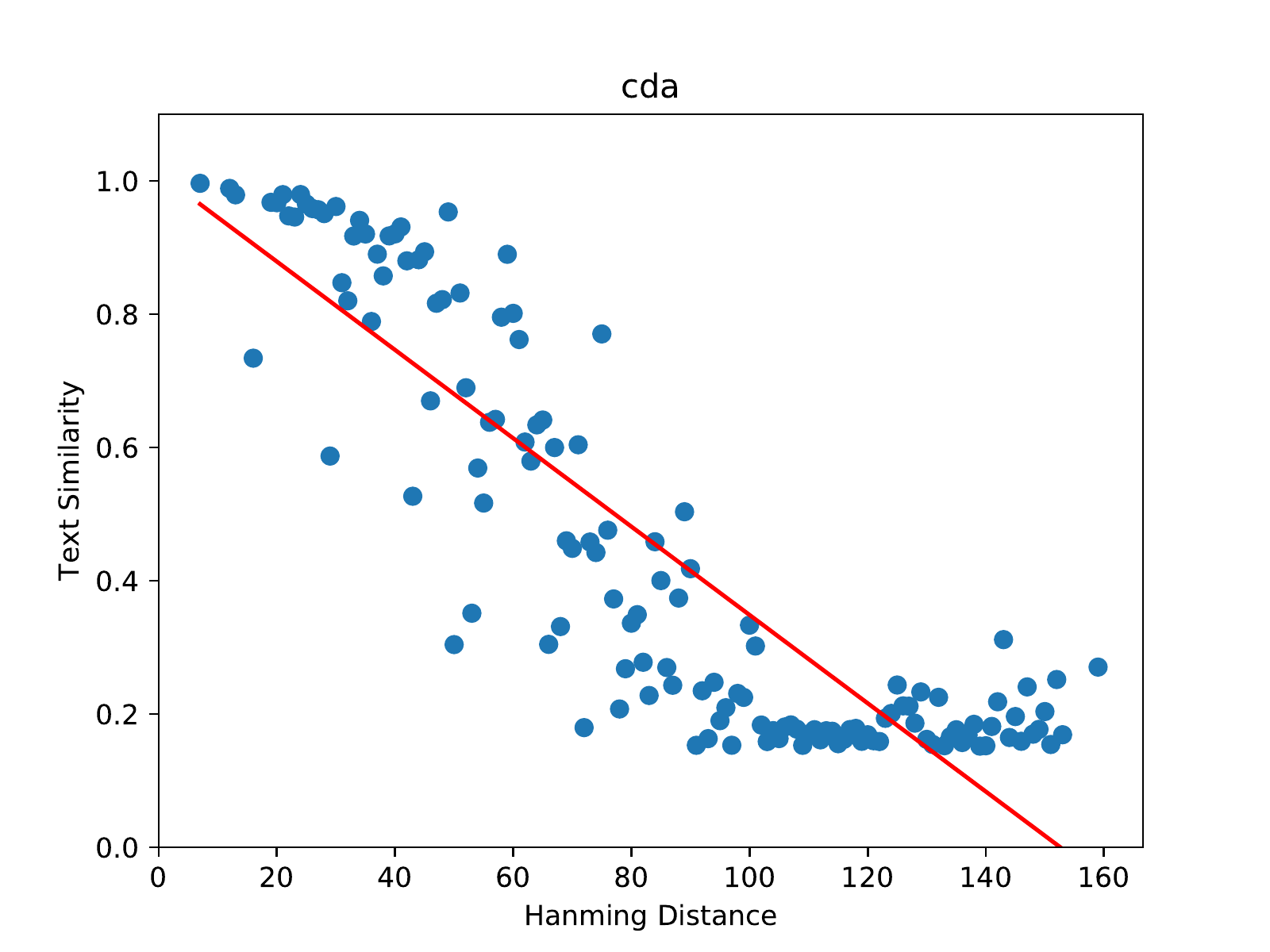}
  }
  \subfigure[douban]{
    %\label{fig:subfig:b} %% label for second subfigure
    \includegraphics[width=2.2in]{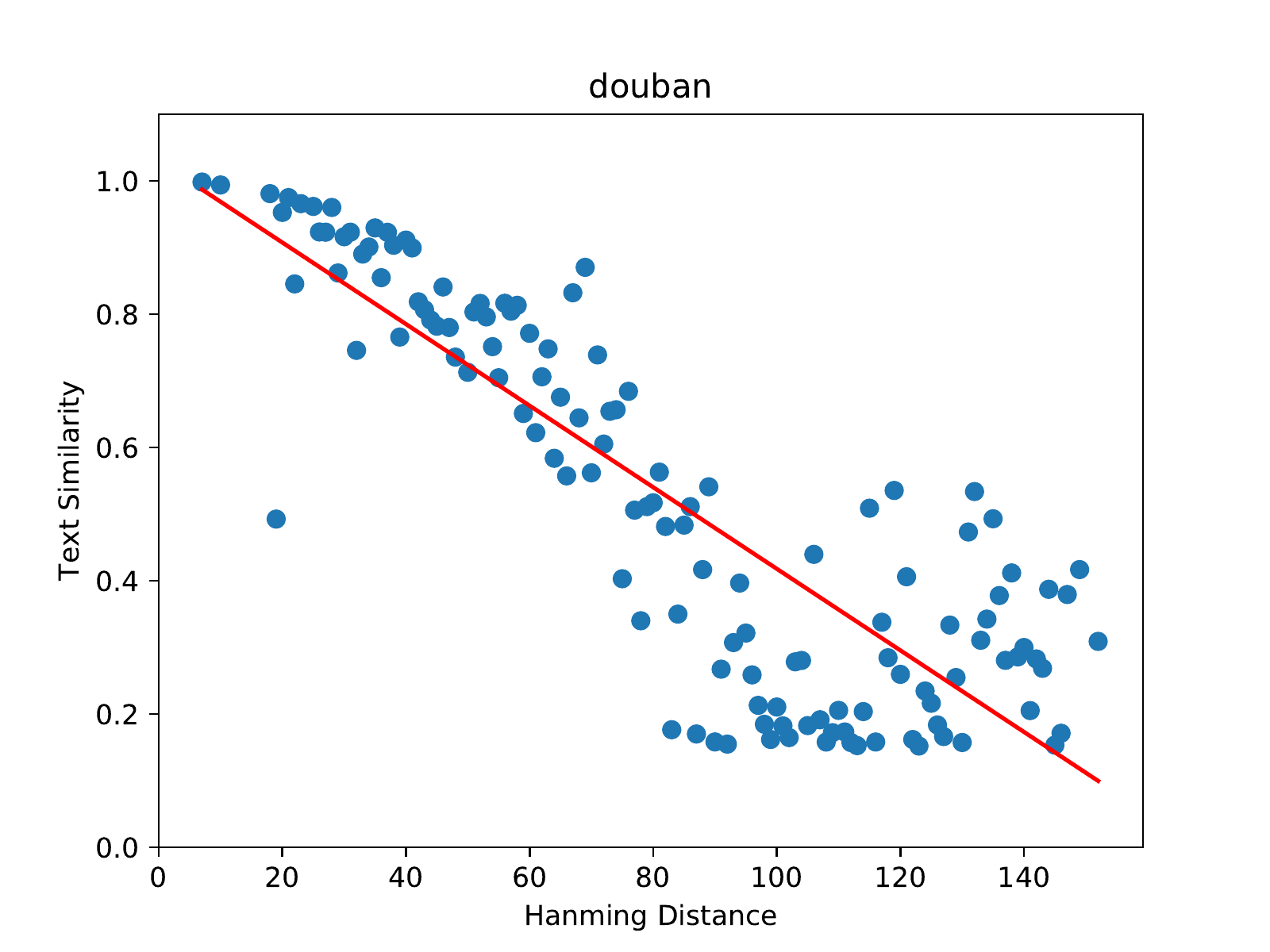}
  }
   \subfigure[sougou news]{
    %\label{fig:subfig:a} %% label for first subfigure
    \includegraphics[width=2.2in]{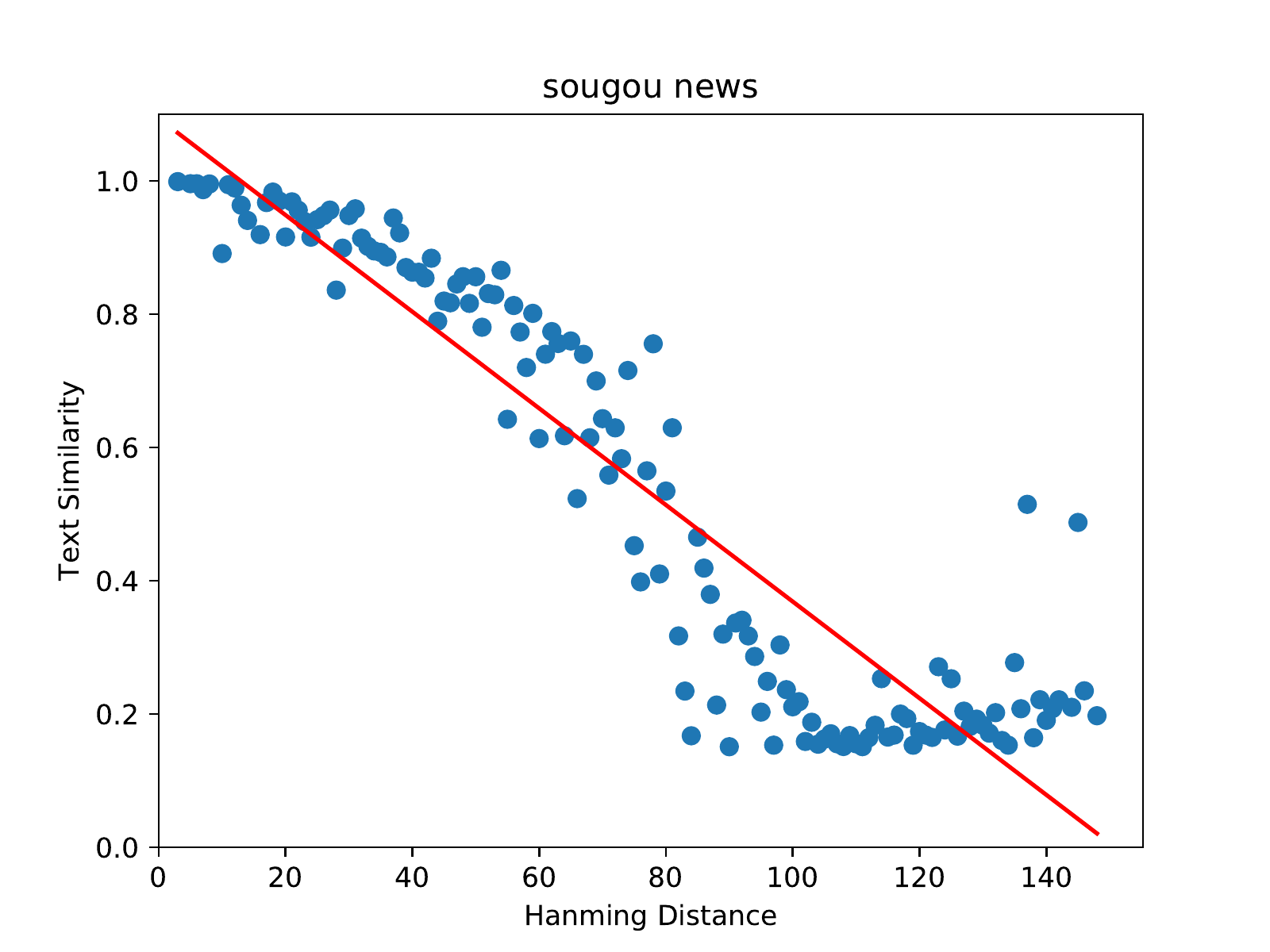}
  }
  \subfigure[weibo]{
    %\label{fig:subfig:b} %% label for second subfigure
    \includegraphics[width=2.2in]{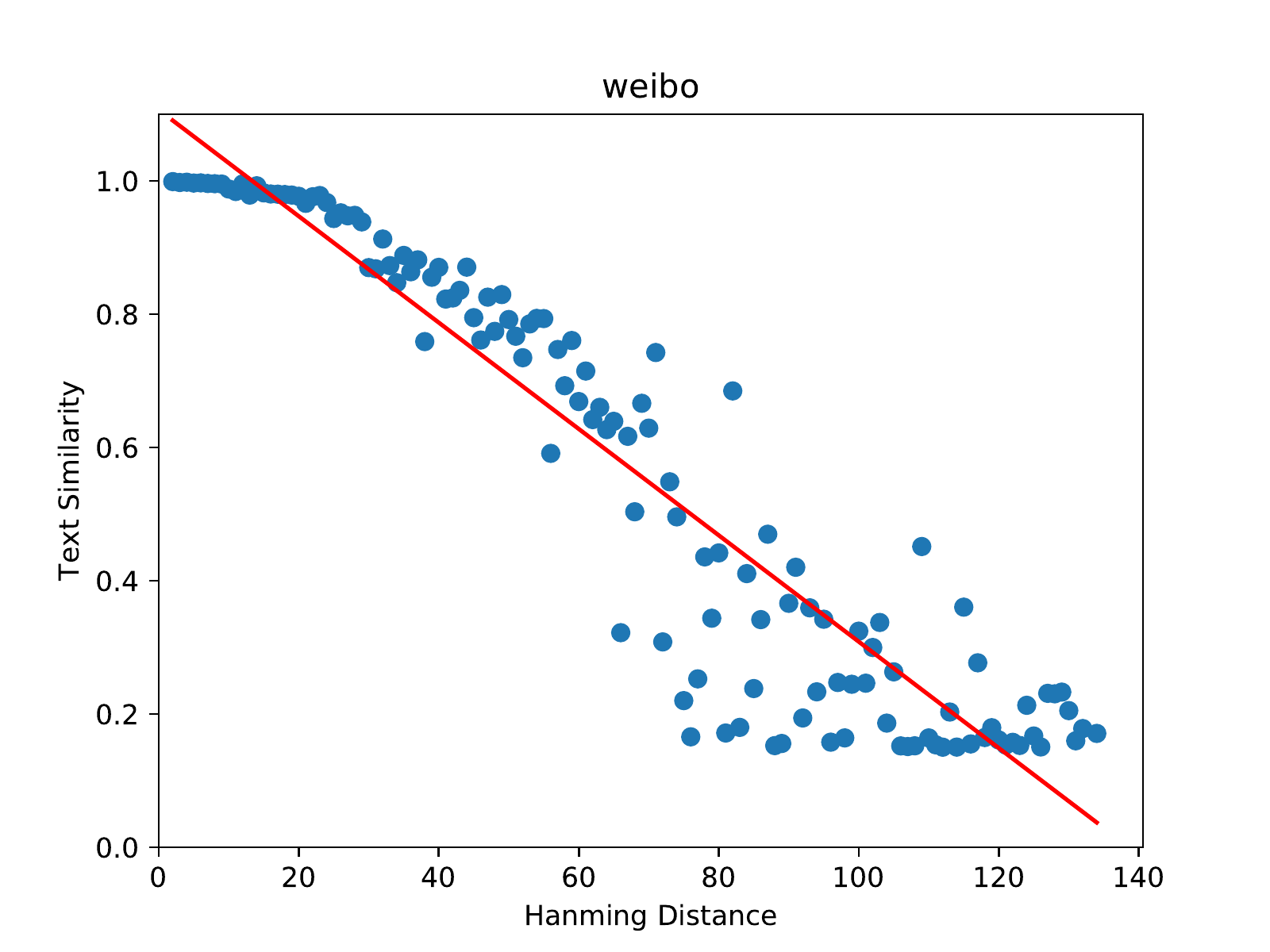}
  }
  \caption{The relationship between Hanming Distances and Text Similarities}
  %\label{fig:subfig} %% label for entire figure
\label{fig:label2}
\end{figure}

We elaborate on the relationship between hamming distances and text similarities. Texts are determined to be partially pirated if the hamming distances exceed a threshold $\theta$.
We conduct experiments on multiple corpora and construct regression models. The four corpora selected for the experiments are clinical document architecture (cda), douban multi-round dialogue (douban), sougou news, sina weibo (weibo), and their related results are illustrated in Figure~\ref{fig:label2}.

The cda corpus is a corpus composed of chief complaints and current medical history in a hospital's production environment. The length of the complaint is about 20 characters. The current medical history is about 100 characters and a total of 700,000 electronic medical records. The douban corpus is a public corpus, the corpus is completely segmented, the average text length is about 81 characters, and the corpus is a total of 520,000 dialogue records. The sogou news corpus is the news data generated by 18 channels of domestic, international, sports, social, entertainment, etc. from June to July 2012. The sogou corpus takes only the body part of the news, containing a total of 220,000 pieces of data, and the average text length is about 210 character. The weibo corpus is a total of 410,000 microblogs generated between June and September 2018, with an average text length of approximately 90 characters.
Based on these corpora, we select 1000 paragraphs of text, and randomly select 500 paragraphs of text for data enhancement to form new text. 1500 paragraphs of text are the sample point of our experiment. The specific data enhancement method is a random addition, deletion, and modification of the text.

The X axis of the four figures represents the hamming distance and the Y axis represents the text similarity in Figure~\ref{fig:label2}. We can see that both hamming distances and text similarities satisfy roughly a linear relationship, which means that we can predict text similaries by evaluating the corresponding hamming distances. The red line is a linear regression model constructed from sample points.
According to the linear regression model, we can predict the corresponding similarity given the hamming distance. Moreover, given a threshold of hamming distance, we can predict text piracy.

\subsubsection{5.1.2 Validation and arbitration Section}

To facilitate the demonstration, we write a smart contract $\emph{SC}$ that meets validation and arbitration requirements. React project was provided by the truffle framework, which assists us to accomplish the design of $\emph{SC}$ on the Rinkey testnet of the Ethereum. The block information of $\emph{SC}$ is shown in Table~\ref{table2}. $\emph{SC}$ will form a special transaction when it is deployed successfully. $\emph{SC}$ contains two functions: $\operatorname{hashIDJudge}()$ and $\operatorname{lshvJudge}()$, and they will form the transactions once they are executed. These transactions have transaction number $\emph{transactionHash}$, transaction index $\emph{txIndex}$, block number $\emph{blockNumber}$ and consumed gas value $\emph{gasUsed}$. The $\operatorname{hashIDJudge}()$ determines whether $\emph{hashID} = \emph{hashID}_{j}$. The $\operatorname{lshvJudge}()$ determines whether $L(\emph{lshv},\emph{lshv}_{j})<=\theta$. $SC$ transfers Ether to the account with correct value if conditions are met.

\begin{table}
\centering
\caption{The block information of $SC$ on the Rinkeby testnet}
\begin{tabular}{cc}

\hline
URL query: & https://rinkeby.etherscan.io/tx/\\

\textbf{Smart contract}&\\
$\emph{transactionHash}$: &\\
\multicolumn{2}{c}{"0xc8867b49c11bf52d4a065c5daefadb9f79ac686}\\
\multicolumn{2}{c}{6fbb48feab5ef252abdaa0859"}\\
$\emph{txIndex}$:& 7 \\
$\emph{blockNumber}$: & 4183700 \\
$\emph{gasUsed}$: & 510710 \\
\hline 

\textbf{hashIDJudge()}&\\
$\emph{transactionHash}$: &\\
\multicolumn{2}{c}{"0x4727abeb77fda42a2d0dd31c6e9fceef5b89b06}\\
\multicolumn{2}{c}{dab3e251d893f54c8293484cf"}\\
$\emph{txIndex}$:& 8 \\
$\emph{blockNumber}$: & 4183728 \\
$\emph{gasUsed}$: & 32650 \\
\hline 

\textbf{lshvJudge()}&\\
$\emph{transactionHash}$: &\\
\multicolumn{2}{c}{"0xca6b55091747a04270e985e8115bb7b4207dbd4}\\
\multicolumn{2}{c}{d6e518d3ed3481d86e803fad4"}\\
$\emph{txIndex}$:& 3\\
$\emph{blockNumber}$:& 4183734  \\
$\emph{gasUsed}$: & 33363 \\

\hline
\label{table2}
\end{tabular}
\end{table}

\subsection{5.2 Performance and Overhead}

We implement the scheme in Local Ganache and Rinkeby Testnet. Our aim is to study whether the detection process can be implemented and compare the time required on the local and test chain. In default, we set only one detection task.

\begin{figure}
  \centering
  \subfigure[Local Ganache]{
    %\label{fig:subfig:a} %% label for first subfigure
    \includegraphics[width=2.2in]{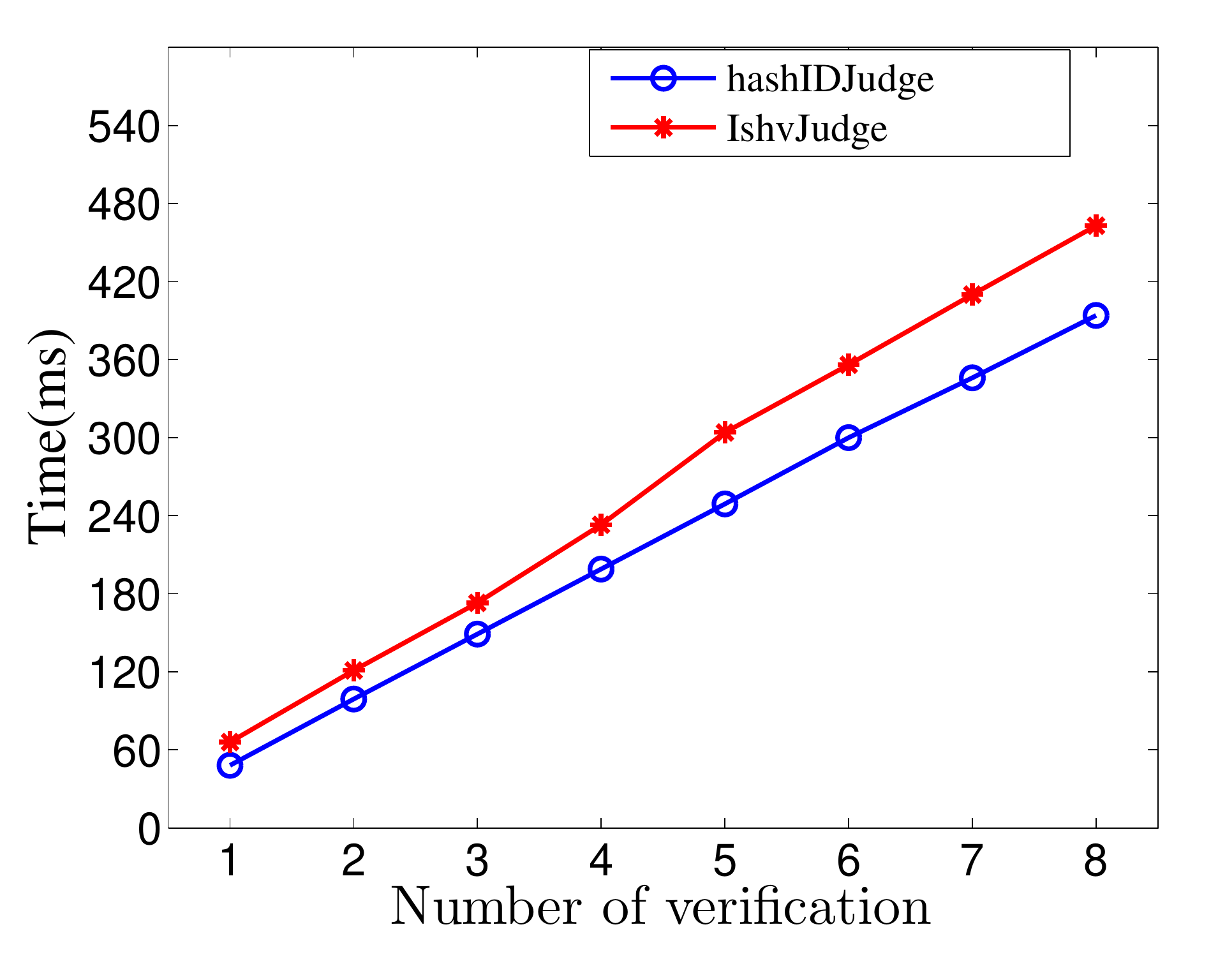}
  }
  \subfigure[Rinkeby Testnet]{
    %\label{fig:subfig:b} %% label for second subfigure
    \includegraphics[width=2.2in]{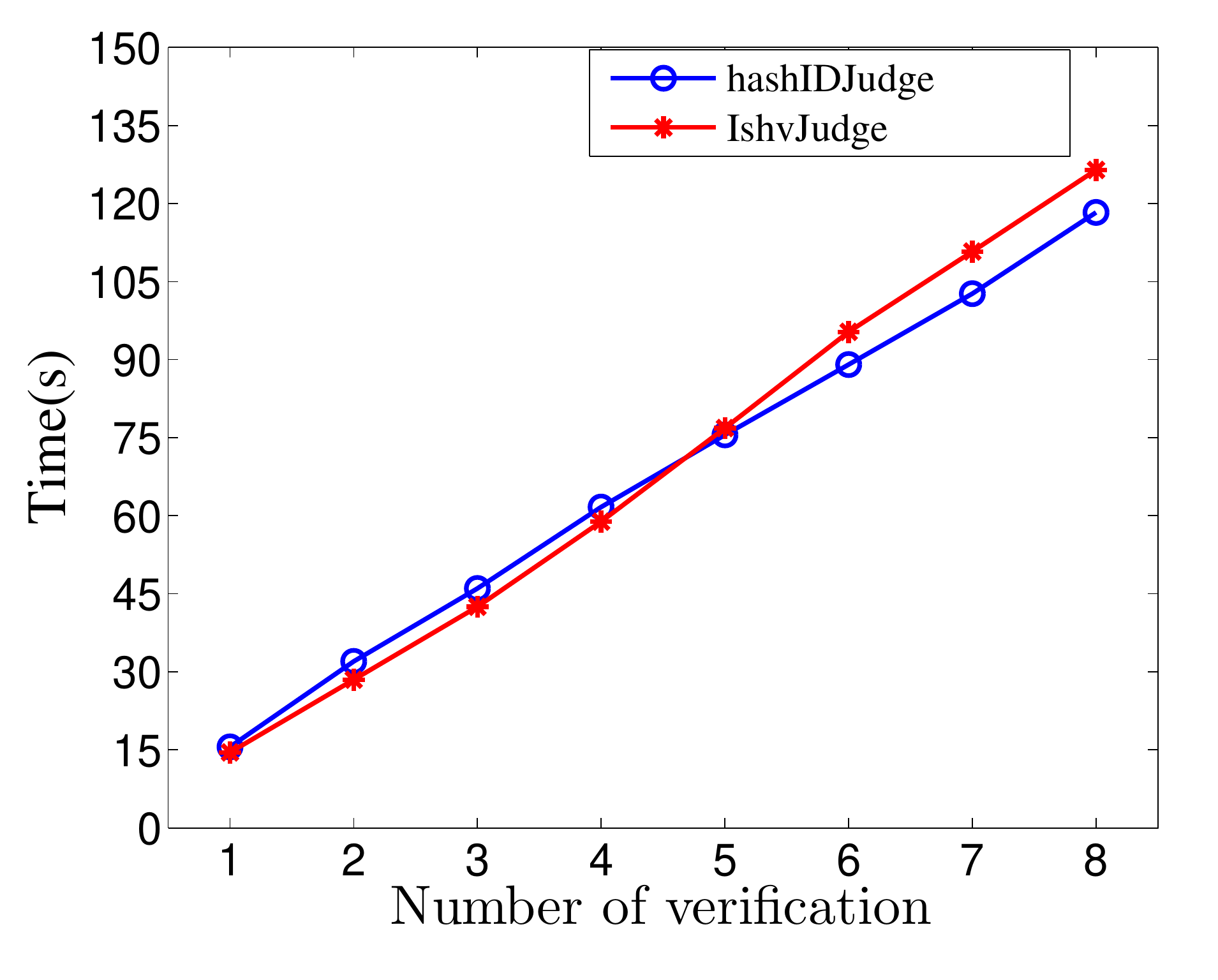}
  }
  \caption{Execution time on local Ganache and Rinkeby testnet}
  %\label{fig:subfig} %% label for entire figure
  \label{fig:label3}
\end{figure}

We observe the efficiency of arbitration functions on the local Ganache and the Rinkeby testnet. As shown in Figure~\ref{fig:label3}, the X axis indicates the number of arbitrations, which are initiated by a single user($\emph{MP}_{i}$). The Y axis indicates execution time. The blue line represents $\operatorname{hashIDJudge}()$ and the red line represents $\operatorname{lshvJudge}()$.
In Figure~\ref{fig:label3}(a), execution time that $SC$ arbitrate one result fluctuates around 60 ms. In Figure~\ref{fig:label3}(b), the execution time of $SC$ arbitrate one result on the Rinkeby testnet fluctuates around 15s. The execution time on the Rinkeby testnet was significantly higher than on the local Ganache.
For the different detection results, the execution time of the validation function is slightly different. In general, the time varies linearly with the number. As the number increases, so does the execution time. Only when users are honest in their detection or verification of copyright can they quickly receive financial rewards.

\section{4. DISCUSSIONs}\label{sec:concludes}

In this paper, we design a copyright detection architecture, DCDChain, using blockchain, SHA256, LSH, and IPFS. DCDChain is applicable to various digital media. The blockchain is used to store the address hash values of legal media and the IPFS is used to store the complete media. The SHA256 and the LSH are applied in the complete and partial piracy detections, respectively. Via DCDChain, the media providers can obtain credible detection results and the detection agent can obtain reasonable service fees. The future work is to design and improve the locality sensitive hash algorithms for various digital media.

\bibliographystyle{oae}
\bibliography{mybibfile}

\end{document}